%% file: Ignatev_2.tex
\documentclass[fleqn,11pt]{article}
\usepackage{amsmath,amssymb}
\usepackage{xcolor}
\usepackage{verbatim}

\def\stackunder#1#2{\mathrel{\mathop{#2}\limits_{#1}}}

\newcommand{\MI}{$\mathfrak{M}_1$}

\newcommand{\MIO}{$\mathfrak{M}^c_1$}

\newcommand{\Fig}[3]{%
\begin{center}
\parbox{8cm}{%
\refstepcounter{figure}\includegraphics[width=8cm,height=#2cm]{#1} \noindent Fig. \thefigure:\quad
#3}\end{center}}
\newcounter{strochka}
\newcommand{\stroka}[1]{\refstepcounter{strochka}\par\noindent\textbf{\arabic{strochka}}. \quad \textsl{#1}}
\newcounter{spisok}

\input preamble.tex

\begin{document}
\thispagestyle{empty}
\twocolumn[

\vspace{1cm}

\Title{Single-field model of gravitational-scalar instability. I. Evolution of perturbations.\foom 1}

\Author{Yu. G. Ignat'ev}
    {Institute of Physics, Kazan Federal University, Kremlyovskaya str., 16A, Kazan, 420008, Russia}


\Abstract
 {On the basis of the previously formulated mathematical model of a statistical system with a scalar interaction of fermions and the theory of gravitational-scalar instability of a cosmological model based on a two-component statistical system of scalarly charged degenerate fermions, a numerical model of the cosmological evolution of gravitational-scalar perturbations for a one-component cosmological system with a canonical scalar interaction is constructed and studied. The influence of the magnitude of the scalar charge of fermions on the differential and integral parameters of the instability is revealed. It is shown that the gravitational-scalar instability in the early stages of expansion in the model under study arises at sufficiently small scalar charges. 4 fundamentally different types of perturbations are identified, as well as 4 types of gravitational-scalar instability, determined by the fundamental parameters of the model. Examples of numerical models are given that provide large values of the increments of the increase in the amplitude of perturbations.
}
\bigskip

] 
\section{Intoduction}
Currently, there is an unsolved problem in astro\-physics and cosmology related to the lack of suffi\-ci\-ently convincing models for the formation of super\-mas\-sive black holes with a mass of the order of $\sim10^9\div10^{10}M_\odot$, which are the central objects of luminous quasars, in the early Universe at $z>6$ and even for $z>7$, which according to modern concepts corresponds to the age of the Universe $0.65\div 1$ billion years (see, for example, \cite{SMBH1}, \cite{SMBH2}, \cite{Fan}). The astrophysical origin of such supermassive black holes in the early Universe is still insufficiently understood, since observational data raise the ques\-tion of the mechanism of formation and rapid growth of such objects in the early Universe.  The results of numerical simulation \cite{Trakhtenbrot} impose a number of restrictions on the formation parameters of super\-mas\-sive black holes. For example, it is shown that light black hole nuclei with a mass $M\leqslant 10^3 M_\odot$ cannot grow to masses of the order of $10^8 M_\odot$ by $z = 6$ even with supercritical accretion. For the formation of supermassive black holes with masses $10^8 \div 10^9 M_\odot$, heavier nuclei are needed by this time
\begin{equation}\label{M_nc}
M_{nc}\sim 10^4\div 10^6 M_\odot
\end{equation}
and gas-rich galaxies containing quasars.  However, at present there are no sufficiently convincing models for the appearance of such heavy nuclei in the early Universe.

Interest in the mechanisms of formation of super\-mas\-sive black holes, taking into account the fact of their dominant presence in the composition of quasars, is caused, in particular, by the fact that such black holes are formed in the composition of quasars at fairly early stages of the evolution of the Universe, before the for\-ma\-tion of stars. This circumstance opens up the possibility of the for\-ma\-tion of super\-mas\-sive black holes under conditions where scalar fields and dark matter can significantly influence this process. Note that the numerical simulation in \cite{Trakhtenbrot} was carried out within the frame\-work of the standard gas accretion model, which does not take into account the possible inf\-lu\-ence of scalar fields on the process of black hole formation. In this connection, we note the papers \cite{Supermass_BH} -- \cite{Soliton}, in which the possibility of the existence of scalar halos and scalar hairs in the vicinity of super\-mas\-sive black holes is considered.

In this regard, it is necessary to construct theo\-re\-tical models for the formation of super\-mas\-sive black holes in the early Universe based on mechanisms different from the standard mechanism of gas accretion. In papers \cite{GC_21_1} -- \cite{GC_21_2} the theory of short-wavelength longitudinal perturbations was formulated in a cosmo\-lo\-gical model based on a degenerate one-component system of scalarly charged fermions. In this case, the fermions were assumed to be singly charged canonical or phantom scalar charge, the potential energy of the corresponding scalar fields was assumed to be Higgs, and the perturbations of the cosmo\-lo\-gical model were studied in the \emph{stiff WKB appro\-xi\-mation}:
\begin{eqnarray}\label{WKB0}
n\eta\gg 1,\\
\label{stiff-WKB}
n^2\gg a^2 m^2,
\end{eqnarray}
where $n$ is the wave number of perturbations in the spatially flat Friedmann metric
\begin{eqnarray}\label{ds0}
ds_0^2=dt^2-a^2(t)(dx^2+dy^2+dz^2)\equiv\nonumber\\
a^2(\eta)(d\eta^2-dx^2-dy^2-dz^2),\;(t=\int ad\eta),
\end{eqnarray}
$m$ is the mass of quanta of the scalar field. In \cite{GC_21_2} the following result was obtained: in the cosmological model based on the canonical scalar field, there always occurs a moment in time at which all short-wavelength longitudinal\footnote{We mean longitudinal perturbations of the gravitational field of the form $\varrho\delta_{\alpha\beta}+\sigma n_\alpha n_\beta$, as well as perturbations of the scalar field corresponding to them.} perturbations become uns\-table, and all perturbations of scalar fields over a finite interval time reach infinitely large values. In this case, the perturbations of the metric remain small. The cosmological model based on phantom interaction turns out to be stable. In \cite{STFI_20} this instability was called \emph{gravitational - scalar ins\-ta\-bi\-lity}\footnote{Although it would be more correct to call it ``scalar - gravitational'', given the growth of perturbations of the scalar field.}.

The stiffness condition of the WKB app\-ro\-xi\-mation \eqref{stiff-WKB} used in \cite{GC_21_2} did not allow us to take into account the specifics of the Higgs potential and study the evolution of perturbations at sufficiently long times. In this regard, in \cite{STFI_20}, based on the complete theory of a two - component system of degenerate scalarly charged fermions with Higgs scalar fields \cite{TMF_Ign_Ign21} and the results of numerical model\-ling of the corresponding cosmological model \cite{GC_21_3}, which will be referred to as the model \MI, a complete theory of the evolution of longitudinal per\-tur\-ba\-tions was constructed with the rigid WKB condition removed - the \eqref{stiff-WKB} app\-ro\-xi\-mation.  First, in \cite{STFI_20}, closed formulas \emph{with respect to a given cosmological background} were obtained for the per\-tur\-ba\-tion eikonal functions in the WKB app\-ro\-xi\-mation \eqref{WKB0} and the assumption of smallness of the fermion scalar charges $e_{(a)}$
\begin{eqnarray}\label{WKB}
n\eta\gg 1;\; n^2\gg  e^4_{(a)},\; a^2m^2_{(a)}\gg e^4_{(a)},
\end{eqnarray}
where $m_{(a)}$ is the mass of quanta of the scalar field $\Phi_{(a)}$\footnote{In the case of $m^2_{(a)}=0$ in \eqref{WKB}, this term is necessary be replaced by $\alpha_{(a)}\Phi^2_{(a)}$, where $\alpha_{(a)}$ is the self-action constant in the Higgs potential of the corresponding field.}. Secondly, the possibility of a short-wave scalar-gravitational instability of the cosmological model in the case of a canonical scalar singlet (canonical field $\Phi$) and an asymmetric scalar doublet (canonical field $\Phi$ and phantom field $\varphi$) was strictly substantiated in this work. , as well as the short-wavelength stability of the cosmological system based on a phantom scalar singlet.

Further, in the work \cite{Ign_GC21_Un} a numerical model of the evolution of scalar-gravitational perturbations was constructed for the case of an asymmetric scalar doublet \MI, examples of the development of instability in a cosmological system were given, and some features of this process were revealed.

As we noted above, it is the canonical scalar field $\Phi$ that is directly responsible for the emergence of instability in the short-wavelength sector of osci\-l\-la\-tions of a degenerate system of scalarly charged fermions. Therefore, the purpose of this paper is, firstly, to construct a numerical model for the evo\-lu\-tion of short-wave perturbations of the \MI\ cosmo\-lo\-gical model for the case of a canonical scalar singlet and a one-component system of degenerate fermions when the condition of the stiff WKB app\-ro\-xi\-mation \eqref{stiff-WKB} is removed. Secondly, the purpose of this work is to study the influence of the model parameters on the process of the emergence and development of disturbance instability and to identify possible features of this process.

The second part of the work will be devoted to the study of the possibility of the formation of black holes in the early Universe using the mechanism of gravitational - scalar instability.

\section{Mathematical model of the cos\-mological system of degenerate scalarly charged fermions}
Bearing in mind the application of a number of results of this article in its second part, we first consider the general mathematical model \MI\ for the case of an asymmetric scalar doublet represented by the canonical scalar field $\Phi$ and a phantom scalar field $\varphi$. The cosmological model for the canonical scalar singlet $\Phi$ will be considered as a special case of the general \MI\ model under the passage of a number of its parameters to the limit.
\subsection{Self-consistent system of equations for degenerate scalarly charged fermions}
Consider a cosmological model based on a self-gravitating two-component system of singly scalarly charged degenerate fermions interacting through a pair of scalar Higgs fields, canonical, $\Phi$, and phantom, $\varphi$. This model is described, firstly, by the system of Einstein equations
\begin{equation}\label{Eq_Einst_G}
R^i_k-\frac{1}{2}\delta^i_k R=8\pi T^i_k+ \delta^i_k \Lambda_0,
\end{equation}
where
\[T^i_k=T^i_{(s)k}+T^i_{(p)k},\]
$T^i_{(s)k}$ is energy-momentum tensor of scalar fields
\begin{eqnarray}\label{T_s}
T^i_{(s)k}=\frac{1}{16\pi }\bigl(2\Phi^{,i}\Phi_{,k} -\delta^i_k\Phi_{,j} \Phi{,j}
 +2V_z(\Phi)\delta^i_k \bigr)\nonumber\\
 -\frac{1}{16\pi }\bigl(2\varphi^{,i}\varphi_{,k} -\delta^i_k\varphi_{,j} \varphi{,j}
 -2V_\zeta(\varphi)\delta^i_k \bigr),
\end{eqnarray}
and
\begin{eqnarray}
\label{Higgs}
V_z(\Phi)=-\frac{\alpha}{4} \left(\Phi^{2} -\frac{m^{2} }{\alpha}\right)^{2};\nonumber\\
V_\zeta(\varphi)=-\frac{\beta}{4} \left(\varphi^{2} -\frac{\mathfrak{m}^{2}}{\beta}\right)^{2}
\end{eqnarray}
are the potential energies of the corresponding scalar fields, $\alpha $ and $\beta$ are their self-action constants, $m$ and $\mathfrak{m}$ are their quantum masses. As a carrier of scalar charges, we consider a two-component degenerate system of fermions, in which the carriers of the canonical charge $z$-fermions have the canonical charge $e_z$ and the Fermi momentum $\pi_{(z)}$, and the carriers of the phantom charge are $\zeta$-fermions have phantom charge $e_\zeta$ and Fermi momentum $\pi_{(\zeta)}$. The dynamic masses of these fermions in the case of zero bare masses are \cite{TMF_Ign_Ign21}
\begin{equation}\label{m_*(pm)}
m_{z}=e_z\Phi;\qquad m_{\zeta}=e_\zeta\varphi.
\end{equation}

The seed cosmological constant $\Lambda_0$, which appears in the right-hand side of the Einstein equations \eqref{Eq_Einst_G}, is related to its observed value $\Lambda$ by the relationship
\[
\Lambda=\Lambda_0-\frac{1}{4}\sum\limits_r \frac{m^4_r}{\alpha_r}.\]

Further, the energy-momentum tensor of an equilibrium statistical system is equal to:
\begin{equation}\label{T_p}
T^i_{(p)k}=(\varepsilon_p+p_p)u^i u_k-\delta^i_k p_p,
\end{equation}
where $u^i$ is the macroscopic velocity vector of the statistical system, $\varepsilon_p$ and $p_p$ are its energy density and pressure. These macroscopic scalars, as well as other scalar functions that determine the macroscopic characteristics of the statistical system, are equal for a two-component statistical system of degenerate fermions (see, for example, \cite{TMF_Ign_Ign21}):
\begin{equation}\label{2_3}
n^{(a)}=\frac{1}{\pi^2}\pi_{(a)}^3;
\end{equation}
\begin{eqnarray}
\label{2_3a_2}
\varepsilon_p=\frac{e^4_z \Phi^4}{8\pi^2}F_2(\psi_z)+\frac{e^4_\zeta \varphi^4}{8\pi^2}F_2(\psi_\zeta);\\
\label{2_3b_2}
p_p  =\displaystyle \frac{e^4_z \Phi^4}{24\pi^2}(F_2(\psi_z)-4F_1(\psi_z))+\nonumber\\
\frac{e^4_\zeta \varphi^4}{24\pi^2}(F_2(\psi_\zeta)-4F_1(\psi_\zeta));\\
\label{2_3c}
\displaystyle
\sigma^z=\frac{e_z^4 \Phi^3}{2\pi^2}F_1(\psi_z);\;\sigma^\zeta=\frac{e_\zeta^4 \varphi^3}{2\pi^2}F_1(\psi_\zeta),
\end{eqnarray}
where the macroscopic scalars $n_{(a)}$ -- \emph{scalar particle number density} and $\sigma^z$ and $\sigma^\zeta$ -- \emph{scalar charge densities} $e_z$ and $e_\zeta$,
\begin{equation}\label{psi_zzeta}
\psi_z=\frac{\pi_{(z)}}{|e_z\Phi|}; \qquad \psi_\zeta=\frac{\pi_{(\zeta)}}{|e_\zeta\varphi|},
\end{equation}
and also to shorten the letter, the functions $F_1(\psi)$ and $F_2(\psi)$ are introduced:
\begin{eqnarray}\label{F_1}
F_1(\psi)=\psi\sqrt{1+\psi^2}-\ln(\psi+\sqrt{1+\psi^2});\\
\label{F_2}
F_2(\psi)=\psi\sqrt{1+\psi^2}(1+2\psi^2)\nonumber\\
-\ln(\psi+\sqrt{1+\psi^2}).
\end{eqnarray}
In addition, we write down the expression we need below for \emph{densities of scalar charges}, $\rho_{(a)}$, which are defined using the charge number density $n_{(a)}$ \cite{Ignat21_TMP} and not coincide in general
with the scalar charge densities $\sigma^z$ and $\sigma^\zeta$ introduced above:
\begin{equation}\label{rho}
\rho_{(a)}=e_{(a)}n_{(a)}=\frac{e_{(a)}\pi^3_{(a)}}{\pi^2}.
\end{equation}
Finally, the equations of scalar fields for the system under study take the form:
\begin{eqnarray}\label{Box(Phi)=sigma_z}
\Box \Phi + m^2\Phi-\alpha\Phi^3 = -\frac{4}{\pi^2}e^4_z\Phi^4 F_1(\psi_z),\\
\label{Box(varphi)=sigma_zeta}
-\Box \varphi + \mathfrak{m}^2\varphi-\beta\varphi^3 = -\frac{4}{\pi^2}e^4_\zeta\varphi^4 F_1(\psi_\zeta).
\end{eqnarray}
Note that the scalar charge densities $\sigma^z$ and $\sigma^\zeta$ \eqref{2_3c} are the sources of the corresponding scalar fields $\Phi$ and $\varphi$, while the scalar charge densities $\rho_z $ and $\rho_\zeta$
are defined by a prime number of charges and, unlike \eqref{2_3c}, are related to the corresponding \emph{conserved scalar charges}
\begin{equation}\label{Q}
Q_{(a)}=\int \rho_{(a)}dV.
\end{equation}
\subsection{Background state for the cosmological model \MI}

Let us further consider the space-flat model of the Friedman universe \eqref{ds0}. A strict consequence of the general relativistic kinetic theory for statistical systems of completely degenerate fermions is the Fermi momentum conservation law $\pi_{(a)}$ for each component
 \begin{equation}\label{ap}
 a(t)\pi_{(a)}(t)=\mathrm{Const}.
 \end{equation}
Assuming in what follows, for definiteness, $a(0)=1$ and
\begin{eqnarray}\label{a-xi}
\xi=\ln a;\quad \xi\in(-\infty,+\infty); \quad \xi(0)=0,\\
\label{psi0}
\pi_{(z)}=\pi_c \mathrm{e}^{-\xi},\; \pi_{(\zeta)}=\pi_f \mathrm{e}^{-\xi},\nonumber\\
 (\pi_c=\pi_{(z)}(0),\pi_f=\pi_{(\zeta)}(0)),
\end{eqnarray}
let us write the complete normal system of Einstein equations and scalar fields $\Phi(t)$ and $\varphi(t)$ for this two-component system of scalarly charged degenerate fermions\cite{TMF_Ign_Ign21} in a clearly nonsingular form:
\begin{eqnarray}
\label{dxi/dt-dPhi_Phi}
\dot{\xi}=H;\qquad \dot{\Phi}=Z;\qquad \dot{\varphi}=z;\\
\label{dH/dt_M1}
\dot{H}=-\frac{Z^2}{2}+\frac{z^2}{2}-\frac{4\mathrm{e}^{-3\xi}}{3\pi}\times\nonumber\\
\biggl(\pi_c^3\sqrt{\pi_c^2\mathrm{e}^{-2\xi}+e^2\Phi^2}+\pi_f^3\sqrt{\pi_f^2\mathrm{e}^{-2\xi}+\epsilon^2\varphi^2}\biggr);
\end{eqnarray}
\begin{eqnarray}
\label{dZ/dt_M1}
\dot{Z}=-3HZ-m^2\Phi+\alpha\Phi^3-\nonumber\\[6pt]
\frac{4e^2\pi_c\mathrm{e}^{-\xi}}{\pi}\Phi\sqrt{\pi^2_c \mathrm{e}^{-2\xi}+e^2\Phi^2}+\nonumber\\
\frac{4e^4}{\pi}\Phi^3\ln\biggl(\frac{\pi_c\mathrm{e}^{-\xi}+\sqrt{\pi^2_c \mathrm{e}^{-2\xi}+e^2\Phi^2}}{|e\Phi|} \biggr);
\end{eqnarray}
\begin{eqnarray}
\label{dzZ/dt_M1}
\dot{z}=-3Hz+\mathfrak{m}^2\varphi-\beta\varphi^3+\nonumber\\[6pt]
\frac{4\epsilon^2\pi_f\mathrm{e}^{-\xi}}{\pi}\varphi\sqrt{\pi^2_f \mathrm{e}^{-2\xi}+\epsilon^2\varphi^2}-\nonumber\\
\frac{4\epsilon^4}{\pi}\varphi^3\ln\biggl(\frac{\pi_f\mathrm{e}^{-\xi}+\sqrt{\pi^2_f \mathrm{e}^{-2\xi}+\epsilon^2\varphi^2}}{|\epsilon\varphi|} \biggr).
\end{eqnarray}
The system of equations \eqref{dxi/dt-dPhi_Phi} -- \eqref{dzZ/dt_M1} has as its first integral the total energy integral \cite{TMF_Ign_Ign21}, which can be used to determine the initial value of the function $H(t)$
\begin{eqnarray}
\frac{Z^2}{2}+\frac{z^2}{2}-\frac{m^2\Phi^2}{2}+\frac{\alpha\Phi^4}{4}-\frac{\mathfrak{m}^2\varphi^2}{2}+\frac{\beta\varphi^4}{4} \nonumber\\
-\frac{e^{-\xi}}{\pi}\biggl(\pi_c\sqrt{\pi_c^2\mathrm{e}^{-2\xi}+e^2\Phi^2}\bigl(2\pi^2_c\mathrm{e}^{-2\xi}+e^2\Phi^2\bigr)\nonumber\\
+\pi_f\sqrt{\pi_f^2\mathrm{e}^{-2\xi}+\epsilon^2\varphi^2}\bigl(2\pi^2_f \mathrm{e}^{-2\xi}+\epsilon^2\varphi^2\bigr)\biggr)\nonumber\\
%
\label{SurfEinst_M1}
+\frac{e^4\Phi^4}{\pi}\ln\biggl(\frac{\pi_c\mathrm{e}^{-\xi}+\sqrt{\pi^2_c \mathrm{e}^{-2\xi}+e^2\Phi^2}}{|e\Phi|} \biggr)\nonumber\\
+\frac{\epsilon^4\varphi^4}{\pi}\ln\biggl(\frac{\pi_f\mathrm{e}^{-\xi}+\sqrt{\pi^2_f \mathrm{e}^{-2\xi}+\epsilon^2\varphi^2}}{|\epsilon\varphi|} \biggr)\nonumber\\
+3H^2-\Lambda=0.
\end{eqnarray}

The \MI\ models were studied in \cite{GC_21_3}, where it is shown that such cosmological models, firstly, have an initial singularity with a short-term ultrarelativistic expansion phase passing to the inflation mode and, secondly, depending on the value cosmological constant can also have a finite singularity.

We introduce the invariant characteristics of the unperturbed cosmological model, which are necessary in what follows, \emph{invariant cosmological acceleration}
\begin{equation}\label{Omega}
\Omega=1+\frac{\dot{H}}{H^2}
\end{equation}
and \emph{invariant curvature of four-dimensional space} $K$
\begin{eqnarray}\label{K}
K\equiv\sqrt{R_{ijkl}R^{ijkl}}=H^2\sqrt{6(1+\Omega^2)}\nonumber\\
\equiv \sqrt{6}\sqrt{H^4+\bigl(H^2+\dot{H}\bigr)^2}\geq0.
\end{eqnarray}

\section{Gravitational-scalar instability of the \MI model in the short-wavelength limit for a one - \newline component system}
\subsection{WKB approximation of instability theory}
For the case of a classical scalar Higgs field and a one-component system of singly scalarly charged degenerate fermions, we reformulate the main results of \cite{STFI_20} (see also \cite{Ign_GC21_Un}), in which the evolution of gravitational-scalar perturbations of the \eqref{ds0} metric and scalar fields in the \MI model, for the case of purely longitudinal perturbations of the metric \eqref{ds0} in the form \cite{Land_Field} (for details, see \cite{Lifshitz})
\begin{eqnarray}
\label{metric_pert}
ds^2=ds^2_0-a^2(\eta)h_{\alpha\beta}dx^\alpha dx^\beta,
\end{eqnarray}
where $ds_0$ is the unperturbed spatially flat Friedmann metric \eqref{ds0} in conformally flat form
and, for definiteness, the wave vector is directed along the $Oz$ axis:
\begin{eqnarray}\label{nz1}
 h_{11}=h_{22} =\frac{1}{3}[\lambda(t)+\frac{1}{3}\mu(t)]\mathrm{e}^{inz};\nonumber\\
\label{nz13}
h=\mu(t)\mathrm{e}^{inz};\; h_{12}=h_{13}= h_{23}=0;\nonumber\\
\label{nz2}
h_{33}=\frac{1}{3}[-2\lambda(t)+\mu(t)]\mathrm{e}^{inz}.
\end{eqnarray}
At the same time, the matter in the \MI model in the case of a classical scalar Higgs singlet and a one-component system of degenerate scalarly charged fermions is completely determined by two scalar functions - $\Phi(z,\eta)$ and $\pi_{(z)}(z,\ eta)$, as well as the velocity vector $u^i(z,\eta)$. Let us expand these functions into a series in terms of the smallness of perturbations with respect to the corresponding functions against the background of the Friedmann metric \eqref{ds0}:\footnote{For scalar singlets, see \cite{GC_21_1}. To avoid cumbersome notation, we have retained the notation for the perturbed values of the functions, distinguishing them only by arguments.}
\begin{eqnarray}
\Phi(z,\eta)=\Phi(\eta)+\delta\Phi(\eta)\mathrm{e}^{inz};\nonumber\\
\label{dF-drho-du}
\pi_{(z)}(z,t)=\pi_{(z)}(\eta)(1+\delta(\eta)\mathrm{e}^{inz});\nonumber\\
\sigma^z(z,\eta)= \sigma^z(\eta)+\delta\sigma^z(\eta)\mathrm{e}^{inz};\\
u^i=\frac{1}{a}\delta^i_4+\delta^i_3 v(\eta)\mathrm{e}^{inz},\nonumber
\end{eqnarray}
where $\delta\Phi(\eta)$, $\delta(\eta)$, $s_z(\eta)$, and $v(\eta)$ are functions of the first order of smallness compared to their unperturbed values.

In \cite{STFI_20} (see also \cite{Ign_GC21_Un}), the evolution of longitudinal gravitational scalar perturbations of the \MI\ model is studied in the short-wavelength and low-charge approximations \eqref{WKB}.
At the same time, in contrast to the works \cite{GC_21_1} -- \cite{GC_21_2}, the condition of the rigid WKB approximation was not imposed in this work, which makes it possible to consider sufficiently large wavelengths:
\[n^2\gtrsim a^2\{ m^2\Phi,\alpha\Phi^3,\mathfrak{m}^2\varphi,\beta\varphi^3\}.\]
In accordance with the WKB method, we represent the perturbation functions $f(\eta)$ in the form
\begin{equation}\label{Eiconal}
f=\tilde{f}(\eta) \cdot \mathrm{e}^{i\int u(\eta)d\eta}; \quad (|u\eta|\sim |n\eta| \gg 1),
\end{equation}
where $\tilde{f}(\eta)$ and $u(\eta)$ are functions of the perturbation amplitude and eikonal that vary slightly along with the scale factor.

In this paper, in contrast to the case of a scalar doublet, we will not impose an additional condition for the smallness of the scalar charge used in \cite{STFI_20} to simplify the dispersion equation
\[n^2\gg  e^4_{(a)},\; a^2\{m^2,\mathfrak{m}^2\}\gg e^4_{(a)}.\]
The equations for perturbation amplitudes in the zero WKB approximation \eqref{WKB0} take the form of a linear homogeneous system of algebraic equations ($\nu=\lambda+\mu$)
\begin{eqnarray}\label{WKB0-sys}
\!\!\!
\left[
\begin{array}{ccc}
n^2-u^2+\gamma_{11} & 0 & n^2\gamma_{13}\\
0 & u^2 & 0 \\
\gamma_{31} & 0 & n^2\gamma_{33}-u^2\\
\end{array}
\right]\cdot
\left[
\begin{array}{c}
\delta\Phi\\ \lambda \\ \nu\\
\end{array}
\right]=0,\nonumber
\end{eqnarray}
where the coefficients $\gamma_{\alpha\beta}$ for the case under study are:
\begin{eqnarray}
\label{gamma_ik}
\gamma_{11}\equiv a^2(m^2-3\alpha\Phi^2+8\pi S^z_\Phi);\nonumber\\
\gamma_{13}\equiv \frac{e^4_z\Phi^3\psi^2_z}{6\pi^2\varepsilon^\delta_p\sqrt{1+\psi^2_z}};\; \gamma_{33}\equiv \frac{1}{3}+\frac{p^\delta_p}{\varepsilon^\delta_p};\\
\gamma_{31}\equiv -3a^2[\Phi(m^2-\alpha\Phi^2)-8\pi P^\Phi].\nonumber
\end{eqnarray}
Coefficients of the theory of gravitational-scalar instability \cite{STFI_20} included in the formulas \eqref{gamma_ik} and expressed in terms of the basic functions of the unperturbed model \MI\ $a(t)$ and $\Phi(t)$, as well as through the kinetic coefficients $\psi_a(t)$ \eqref{psi_zzeta}, are:
\begin{eqnarray}\label{de_delta}
\varepsilon_p^\delta=\frac{1}{\pi^2}e_z^4\Phi^4\psi_z^3\sqrt{1+\psi_z^2}>0;\nonumber\\
\varepsilon_p^\Phi=\frac{e_z^4\Phi^3}{2\pi^2}F_1(\psi_z);
\quad \Delta_\Phi=\frac{\varepsilon_p^\Phi}{ 8\pi \varepsilon_p^\delta};\\
%
\!\!\!p_p^\delta=\frac{1}{\pi^2}\frac{e_z^4\Phi^4\psi_z^4}{\sqrt{1+\psi_z^2}}
>0;\hskip 0.5mm\nonumber\\
%
\label{P^Phi}
\!\!\!P^\Phi=\frac{e^4_z\Phi^3}{2\pi^2}F_1(\psi_z)-p_p^\delta\Delta_\Phi;\\
\label{S^z_Phi}
S^z_\Phi=\frac{e^4_z\Phi^2}{2\pi^2}\biggl(3F_1(\psi_z)\nonumber\\
 -\frac{\psi^3_z}{\sqrt{1+\psi^2_z}}-\frac{\psi^2_z}{\sqrt{1+\psi^2_z}}\Delta_\Phi\biggr).
\end{eqnarray}
Using these coefficients, in turn, macroscopic scalars are defined:
\begin{eqnarray}\label{s_z-delta}
\!\!\!\delta\sigma^z=\frac{e^4_z\Phi^3\psi^2_z}{48\pi^3 a^2\varepsilon^\delta_p\sqrt{1+\psi^2_z}}n^2\nu +S^z_\Phi\delta\Phi;&\nonumber\\
\label{dp-delta}
\delta p_p=\frac{p^\delta_p n^2}{24\pi^3a^2\varepsilon^\delta_p}\nu +P^\Phi\delta\Phi. &\nonumber
\end{eqnarray}

\subsection{Dispersion Equation, Modes, and Types of Perturbations \label{disp_eq}}
A necessary and sufficient condition for the nontrivial solvability of the system of equations \eqref{WKB0-sys} is that the determinant of the matrix of this system is equal to zero, which gives the necessary \emph{dispersion equation} on the eikonal function $u(t)$ of perturbations. In this case, two zero modes $u^\pm_{(0)}=0$ are immediately distinguished, corresponding to perturbations of the $\lambda$ metric (see \cite{Ign_GC21_Un}), which are eliminated by admissible transformations\footnote{For details, see \ cite{Lifshitz}}. The four oscillation modes $u^\pm_{(\pm)}$ corresponding to perturbations of the classical scalar field $\delta\Phi$ and perturbations of the $\nu$ metric are determined by the dispersion equation
\begin{eqnarray}\label{A_gamma=0}
\!\!\mathrm{Det}(\bar{\mathbf{A}})=\left|\begin{array}{ll}
n^2-u^2+\gamma_{11}  & n^2\gamma_{13}\\[12pt]
\gamma_{31} &  n^2\gamma_{33}-u^2\\
\end{array}\right|=0,
\end{eqnarray}
The dispersion equation \eqref{A_gamma=0} is a biquadratic equation with respect to the eikonal function $u(t)$, solving which we find solutions:
\begin{eqnarray}
\label{u_2_pm}
u^2_\pm=\frac{1}{2}(n^2(1+\gamma_{33})+\gamma_{11}\nonumber\\
\pm\frac{1}{2}\sqrt{[n^2(1-\gamma_{33})+\gamma_{11}]^2+4\gamma_{13}\gamma_{31}};
\end{eqnarray}
\begin{eqnarray}\label{u_pm}
\Rightarrow u^\pm_\pm=\pm \sqrt{\frac{c\pm \sqrt{b}}{2}}:\hspace{3.05cm}\\
u^+_+=+\sqrt{\frac{c+\sqrt{b}}{2}};\; u^-_+=-\sqrt{\frac{c+\sqrt{b}}{2}};\nonumber\\
u^+_-=+\sqrt{\frac{c-\sqrt{b}}{2}};\; u^-_-=-\sqrt{\frac{c-\sqrt{b}}{2}},\nonumber
\end{eqnarray}
where the upper signs correspond to the signs before the external radical, the lower ones correspond to the sign before the internal one, and the notation
\begin{eqnarray}\label{u_pm_a}
c=n^2+\gamma_{11}+n^2\gamma_{33};\\ b=[n^2(1-\gamma_{33})+\gamma_{11}]^2+4\gamma_{13}\gamma_{31}.\nonumber
\end{eqnarray}
In this case, the relation
\begin{equation}\label{a^2-b}
\frac{1}{4}(c^2-b)\equiv n^2\gamma_{33}(n^2+\gamma_{11})-\gamma_{13}\gamma_{31}.
\end{equation}
\eqref{u_pm} solutions satisfy the following relations:
\begin{eqnarray}\label{u-=-u+}
u^-_\pm=-u^+_\pm;\\
u^+_-u^+_+=u^-_-u^-_+=\sqrt{c^2-b},\nonumber
\end{eqnarray}

Accordingly, \eqref{u_pm}, there are only 4 types of perturbations, depending on the ratio between the quantities $a$ and $b$:
\begin{eqnarray}
\label{b>0,b>a^2}
\mathbf{1.}\; b>c^2:\Rightarrow \hspace{5cm} \\
\Im(u^+_+)=\Im{u^-_+}=0;\; \Re(u^+_+)=-\Re(u^-_+);\nonumber\\
\Re(u^+_-)=\Re(u^-_-)=0;\; \Im(u^+_-)=-\Im(u^-_-)\nonumber
\end{eqnarray}
\begin{eqnarray}
\label{a^2>b>0}
\mathbf{2.} & c>0,\ 0<b<c^2: \Rightarrow \Im(u^\pm_\pm)=0;\\
 & \Re(u^+_+)=-\Re(u^-_+); \Re(u^-_+)=-\Re(u^-_-);\nonumber
\end{eqnarray}
\begin{eqnarray}
\label{b<0}
\mathbf{3.} & b<0: \Rightarrow (u^+_+)^*=u^+_-;\; (u^-_+)^*=u^-_-;
\end{eqnarray}
\begin{eqnarray}
\label{b<a^2,a<0}
\mathbf{4.} & c<0,\ 0<b<c^2: \Rightarrow \Re(u^\pm_\pm)=0;\\
 & \Im(u^+_+)=-\Im(u^+_-);\;  \Im(u^-_+)=-\Im(u^-_-); \nonumber
\end{eqnarray}

Of the listed 4 types of perturbations \eqref{b>0,b>a^2} -- \eqref{b<a^2,a<0} 1st \eqref{b>0,b>a^2} the type represents a superposition of a pair of standing growing and damping modes and a pair of undamped modes\footnote{Standard weak amplitude decay occurs only due to the geometric factor $a(t)$.} of retarded and advanced waves, 2nd \eqref{a^2>b>0} type represents two pairs of undamped waves (leading and retarded) with different frequencies, 3rd \eqref{b<0} type - two pairs of traveling waves (leading and retarded) with different frequencies, having damped and growing modes, 4 -th \eqref{b<a^2,a<0} type represents pairs of damped and growing over time with different decrements/inc\-re\-ments of standing waves. Since, according to \eqref{u_pm_a}, the coefficients $c$ and $b$ are functions of time -- $c(t),\ b(t)$, the listed types of perturbations can transform one into another at times $t_k$:
\begin{eqnarray}\label{perehod}
b(t)>0\rightarrow b(t_k)=0\rightarrow b(t)<0; \nonumber\\
c(t)>0\rightarrow c(t_k)=0\rightarrow c(t)<0; \nonumber\\
c^2(t)-b(t)>0 \rightarrow \nonumber\\
c(t_k)^2-b(t_k)=0\rightarrow c(t)^2-b(t)<0.
\end{eqnarray}
As a result, in the course of cosmological evolution, a rather complex picture of the alternation of the stages of wave oscillations and the stages of in\-s\-ta\-bi\-lity, determined by the presence of the imaginary part
of the eikonal, can be obtained.

\subsection{Frequencies and decrement\!\! /\!\! decay increment\!\! /\!\! oscillation instability}
Further, due to the linearity of the perturbations, the final expressions for the perturbations according to the formula \eqref{Eiconal} and the found values of the eikonal functions \eqref{u_pm} can be written in the form:
\begin{eqnarray}\label{Sol_perts}
f=\mathrm{e}^{inz}\sum\limits_\pm\sum\limits_\pm\tilde{f}^{\pm}_{\pm}(\eta)\mathrm{e}^{i\int u^\pm_\pm(\eta)d\eta}+\mathbf{CC},
\end{eqnarray}
where $\tilde{f}^{\pm}_{\pm}(\eta)$ are slowly varying perturbation am\-pli\-tudes $\nu(\eta,z)$ and $\delta\Phi(\eta,z) $ corresponding to the above four oscillation modes $u^\pm_{\pm}$, $\mathbf{CC}$ means the complex conjugate quantity. Thus, unstable modes can only correspond to eikonal functions $u^\pm_\pm(\eta)$ in the region of their complex values.

Passing to the cosmological time using the for\-mula \eqref{ds0} in the expressions \eqref{Sol_perts} and separating the real and imaginary parts in the eikonal functions, we obtain
\begin{eqnarray}
i\int\limits_{\eta_0}^\eta u^\pm_\pm d\eta=i\int\limits_0^t \frac{u^\pm_\pm(t)}{a(t)}dt\nonumber\\
\equiv i\int\limits_0^t \omega^\pm_\pm dt-\int\limits_0^t \gamma^\pm_\pm dt,
\end{eqnarray}
where $\omega(t)$ and $\gamma(t)$ are local frequency and de\-c\-rement\!\! /\!\! decay increment\!\! /\!\! instability of oscillations on the scale of cosmological time $t$:
\begin{eqnarray}\label{g,o}
\omega^\pm_\pm=\mathrm{e}^{-\xi(t)}\mathrm{Re}\left(u^\pm_\pm\right)\equiv \mathrm{e}^{-\xi(t)}\tilde{\omega}^\pm_\pm;\nonumber\\
\gamma^\pm_\pm=-\mathrm{e}^{-\xi(t)}\mathrm{Im}\left(u^\pm_\pm\right)\equiv \mathrm{e}^{-\xi(t)}\tilde{\gamma}^\pm_\pm,
\end{eqnarray}
and $\tilde{\omega}^\pm_\pm$ and $\tilde{\gamma}^\pm_\pm$ are local frequency and decrement\!\! /\!\! decay increment\!\! /\!\! increase in fluctuations on the scale of the time variable $\eta$.

The parity of the eikonal \eqref{u-=-u+} implies the parity property of the oscillation frequency and increment\footnote{If there are corresponding real and imaginary parts of the eikonal function}
\begin{equation}\label{omega-=-omega+}
\omega^+_\pm=-\omega^-_\pm\equiv \omega_\pm;\; \gamma^+_\pm=-\gamma^-_\pm\equiv \gamma_\pm.
\end{equation}
Therefore, the terms under the double sum \eqref{Sol_perts} corresponding to each pair of modes can be written in the form
\[\displaystyle \tilde{f}^+_\pm \mathrm{e}^{i(nz+\int \omega_\pm dt)}\mathrm{e}^{-\int\gamma_\pm dt} +\tilde{f}^-_\pm \mathrm{e}^{i(nz-\int \omega_\pm dt)}\mathrm{e}^{+\int\gamma_\pm dt}.\]

Now adding to this expression its complex con\-ju\-gate, we obtain, according to \eqref{Sol_perts}, the final exp\-res\-sion for perturbations
\begin{eqnarray}\label{Sol_perts1}
f= \biggl(\tilde{f}^+_\pm \mathrm{e}^{i(nz+\int \omega_\pm dt)}\nonumber\\
+(\tilde{f}^+_\pm)^* \mathrm{e}^{-i(nz+\int \omega_\pm dt)}\biggr) \mathrm{e}^{-\int\gamma_\pm dt}\nonumber\\
+\biggl(\tilde{f}^-_\pm \mathrm{e}^{i(nz-\int \omega_\pm dt)}\\
+(\tilde{f}^-_\pm)^* \mathrm{e}^{-i(nz-\int \omega_\pm dt)}\biggr)\mathrm{e}^{+\int\gamma_\pm dt}.\nonumber
\end{eqnarray}

In the general case, disturbances represent two groups of retarded and advanced waves propagating with a phase velocity
\begin{equation}\label{v_f}
v_f=\frac{\varpi}{n}\equiv a\frac{\omega_\pm}{n}
\end{equation}
with \emph{exponentially} decaying or growing amplitudes
\begin{equation}\label{instability}
\tilde{f}^-(\eta)\mathrm{e}^{ -\int\tilde{\gamma}(n,\eta)d\eta},\; \tilde{f}^+(\eta)\mathrm{e}^{+\int\tilde{\gamma}(n,\eta)d\eta}.
\end{equation}
The growing oscillation modes correspond to the instability of the homogeneous unperturbed state of the cosmological model. As we noted above, this mode is associated with $\{\delta\Phi,\nu\}$ disturbances, so the instability, if it exists, is essentially \emph{gravi\-ta\-tio\-nally - scalar} in nature. Further, according to \eqref{instability}, the amplitude of the growing disturbance mode at time $t$, \emph{growth factor of the disturbance amplitude}, is determined by the expression
\begin{equation}\label{chi}
\chi(t)=\int\limits_{t_1}^t \gamma(t)dt,
\end{equation}
where $t_1$ is the initial moment of instability occu\-rrence. Let $t_2$ be the end time of the unstable phase, so that for $t>t_2$ $\gamma(t)=0$. Thus, during the development of instability on the interval $\Delta t=t_2-t_1$, the perturbation amplitude is fixed at $\tilde{f}^+(t)\exp(\chi_\infty)$, where
\begin{equation}\label{chi8}
\chi_\infty=\int\limits_{t_1}^{t_2} \gamma(t)dt.
\end{equation}
\section{Preliminary remarks on\newline numerical modelling}
\subsection{Parameters and initial conditions}
The general background model \MI\ is defined by an ordered set of 9 parameters \cite{GC_21_3}
\begin{equation}\label{Params}
\mathbf{P}=[[\alpha,m,e,\pi_c],[\beta,\mu,\epsilon,\pi_f],\Lambda]
\end{equation}
and initial conditions
\begin{equation}\label{Inits}
\mathbf{I}=[\Phi_0,Z_0,\varphi_0,z_0,\kappa],
\end{equation}
where $\kappa=\pm1$, and the value $\kappa=+1$ corresponds to the non-negative initial value of the Hubble parameter $H_0=H_+\geqslant0$, and the value $\kappa=-1$ corresponds to the negative initial value of the Hubble parameter $H_0 =H_-<0$.
In this paper, we consider a particular case of the \MI cosmological model based on a one-component degenerate system of fermions charged with a canonical scalar charge for a one-field model of a scalar field. This special case is obtained from the \MI\ model with the following \eqref{Params} parameter values and \eqref{Inits} initial conditions: $\pi_f=0$, $\varphi\to0$, $z=0$:
\begin{equation}\label{ParamsM0}
\mathbf{P}=[[\alpha,m,e,\pi_c],\Lambda],
\end{equation}
\begin{equation}\label{Inits1M0}
\mathbf{I}=[\Phi_0,Z_0,\kappa].
\end{equation}
In what follows, we will denote such a model by the symbol \MIO.

Note, firstly, that according to the equation \eqref{dH/dt_M1} in the absence of a phantom field in the \MI\ model, always
\begin{equation}\label{dotH<=0}
\eqref{dH/dt_M1}\displaystyle \stackunder{\varphi=0}{\Rightarrow} \dot{H}\leqslant 0,
\end{equation}
and the zero value $\dot{H}=0$ can be reached only for $\Phi=\Phi_0$ and $\xi\to+\infty \Rightarrow a(t)\to+\infty$, i.e., in infinite the future. In this case, \eqref{dZ/dt_M1} implies that $\Phi_0$ is determined by a stationary singular point of a dynamical system with a vacuum scalar field (see \cite{Ignat21_TMP}). However, in the general case, the existence of this stationary point may contradict the Einstein equation \eqref{SurfEinst_M1}, which imposes a rigid connection between the possible fundamental parameters of the cosmological model \cite{TMF_Ign_Ign21}.

In what follows, we will need the coordinates of the singular points of the dynamical system of the unperturbed cosmological model with the vacuum canonical Higgs field in the phase plane $[\Phi,H],\ Z=0$. There can be 8 such points under the condition $\alpha>0,\Lambda\geqslant0$ (see \cite{Ignat21_TMP}): 4 symmetrical pairs:
\begin{eqnarray}\label{sing_point}
M^\pm_0=\biggl[0,\pm\sqrt{\frac{\Lambda}{3}}\biggr];\\
M^\pm_{\pm1}=\biggl[\pm \frac{m}{\sqrt{\alpha}},\pm\sqrt{\frac{\Lambda}{3}+\frac{m^4}{12\alpha}}\biggr].\nonumber
\end{eqnarray}
According to the results of \cite{Ignat21_TMP}, adapted to the case of a one-field model, \emph{under the condition that singular points exist} \eqref{sing_point}, they have the following character: the points $M^+_0$ are attracting, the re\-maining points: $M^- _0$; $M^\pm_{\pm1}$ are saddle. As can be seen from \eqref{sing_point}, $M^+_0$ attracting singular points exist only if the cosmological constant $\Lambda\geqslant0$ is non-negative. Due to this
As a result, the behavior of the basic functions of the \MIO\ model critically depends on the sign of the cosmological constant $\Lambda$. In this case, the coordinates of the singular points for the vacuum Higgs scalar field \eqref{sing_point}, of course, \emph{do not depend on the value of the scalar charge}.

Below, using the autonomy of the dynamical system, we set $\xi(0)=0$ everywhere. Thus, the \MI model is determined by 5 fundamental parameters and 3 initial conditions.
Further, keeping in mind the still too large number of parameters and this model, \emph{in this article we will fix some of them, assuming in the future}:
\begin{equation}\label{Params-Inits00}
\mathbf{P}=[[1,1,e,0.1],\Lambda],\;
\mathbf{I}=[1,0,1].
\end{equation}
Thus, the model studied \emph{in this article} has only two parameters $e$ and $\Lambda$. In this case, the coordinates of the singular points of the dynamical system of the unperturbed cosmological model with the vacuum canonical Higgs field \eqref{sing_point} take the following values, depending only on $\Lambda$:
\begin{eqnarray}\label{sing_point00}
\!\!\!M^\pm_0=\biggl[0,\pm\sqrt{\frac{\Lambda}{3}}\biggr];\;
M^\pm_{\pm1}=\biggl[\pm 1,\pm\sqrt{\frac{1}{12}+\frac{\Lambda}{3}}\biggr].
\end{eqnarray}
\subsection{Remarks on the mechanism\newline of instability}
In \cite{Ign_GC21_Un} and \cite{STFI_20}, as well as earlier in \cite{GC_21_2}, it is noted that the gravitational-scalar instability of short-wavelength perturbations in a system of scalarly charged particles develops due to the canonical scalar field, as long as it is still true strict WKB condition. Below are plots of the dependence of the square of the eikonal function $u^2_\pm$ \eqref{u_2_pm} on the value of the wavenumber $n$. \ref{ignatev1} and scale function $\xi$ in the case of a scalar singlet with potential $\Phi=1$. The graphs below in Fig. \ref{ignatev1} -- \ref{ignatev2} are constructed for the following values of fundamental parameters\footnote{Note that the eikonal function $u(t)$ does not explicitly depend on the value of $\Lambda$. its dependence is determined indirectly in terms of the basic functions $a(t)$ and $\Phi(t)$, whose evolution essentially depends on the value of $\Lambda$.}:
\begin{equation}\label{Parsams0}
\mathbf{P_0}=[[1,1,1,1,0.1],\Lambda].
\end{equation}
A necessary and sufficient condition for the onset of instability is
\begin{equation}\label{Im(u)not=0}
\Im(u)\not=0.
\end{equation}
\emph{In particular}, instabilities arise in the region of negative values of the square of the eikonal function
\begin{equation}\label{u2<0}
u^2<0.
\end{equation}
\Fig{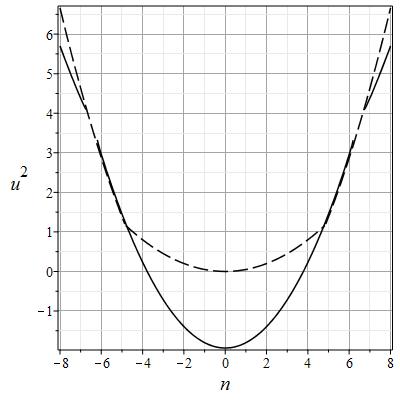}{7}{\label{ignatev1}Dependence of the square of the eikonal function $u^2_\pm$ on the wave number at $\xi=1$ for the \eqref{Parsams0}. The dashed line is $u^2_+$ and the solid line is $u^2_-$.}

We emphasize, firstly, that the condition \eqref{u2<0} is not necessary, but only sufficient, since the square of the eikonal function \eqref{u_2_pm} can also be a complex quantity, provided that the expression under the radical \eqref{u_2_pm}
\begin{equation}\label{uslov:b<0}
b\equiv (n^2+\gamma_{11}-n^2\gamma_{33})^2+4\gamma_{13}\gamma_{31}<0,
\end{equation}
which can be satisfied for sufficiently large values of the scalar charge $e$. In this case, all functions of the eikonal $u^\pm_\pm(t)$ automatically become complex, which ensures the occurrence of instability (3rd type of perturbations \eqref{b<0}). This case just cor\-res\-ponds to Fig. \ref{ignatev2}.

Next, in Fig. \ref{ignatev3} -- \ref{ignatev4} shows the dependence of the oscillation frequency $\omega=\Re(u_-)$ and the increment\!\! /\!\! decrement of growth\!\! /\!\! the damping of the perturbation amplitude $\gamma=-\Im(u_-)$ on the value of the scale function $\xi$.
\Fig{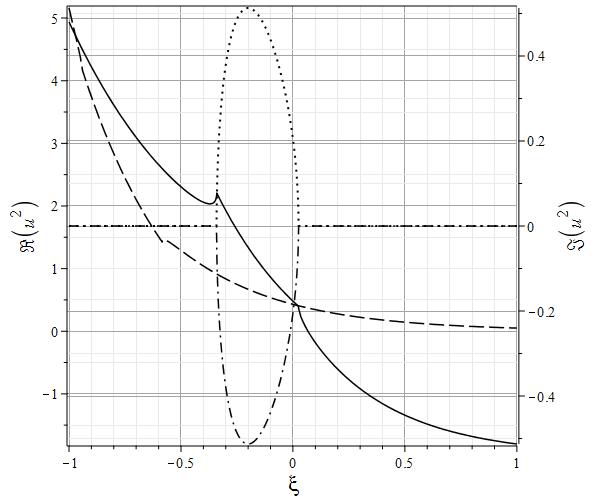}{7}{\label{ignatev2}Dependence of the real and imaginary parts of the square of the eikonal function $u^2_\pm$ on the scaling function $\xi$ for $n=1$ for the parameters \eqref{Parsams0}. Solid line -- $\Re(u^2_-)$, dash-dotted line -- $\Re(u^2_+)$, dashed line -- $\Im(u^2_-)$, dotted line -- $\Im(u^2_+)$. }
\Fig{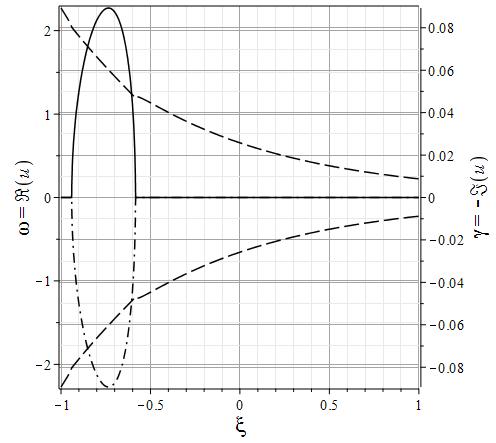}{7}{\label{ignatev3}Dependence of the frequencies $\omega^\pm_+ =\Re(u^\pm_+)$ (dashed lines) and the growth rate of the perturbation amplitude $\gamma=-\Im(u^-_+)$, $\gamma=-\Im(u^+_+)$ (solid and dash-dotted lines, respectively) on the value of the scaling function $\xi$ at $n=1$ for the parameters \eqref{Parsams0}. }

The above examples show the fundamental pos\-si\-bility of the existence of a gravitational-scalar instability in a system of scalarly charged fermions with a classical scalar Higgs interaction.

\Fig{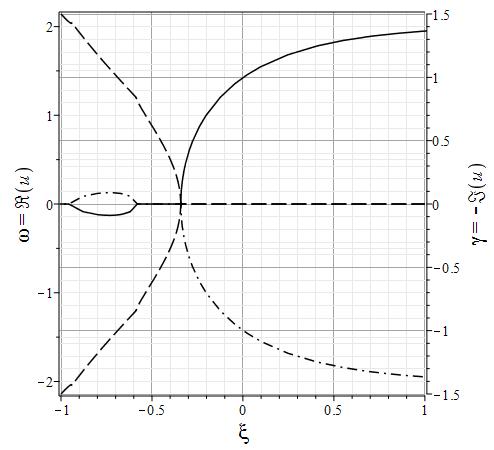}{7}{\label{ignatev4}Dependence of the frequencies $\omega^\pm_- =\Re(u^\pm_-)$ (dashed lines) and the growth rate of the disturbance amplitude $\gamma=-\Im(u^-_-), \ \gamma=-\ Im(u^-_+)$ (solid and dash-dotted lines, respectively) on the value of the scale function $\xi$ at $n=1$ for the parameters \eqref{Parsams0}. }
However, the question of the emergence of ins\-ta\-bi\-lity in the course of cosmological evolution remains open. Indeed, the condition for the occurrence of instability \eqref{Im(u)not=0} is, in essence, an algebraic condition. But for the emergence of a gravitational-scalar instability, it is necessary that in the process of cosmological evolution, on a certain time interval, the values of the basic functions of the model $a(t),\Phi(t)$ turn out to be such that the condition \eqref{Im(u)not=0}
\begin{equation}\label{u2(t,n)<0}
\Im(u(a(t),\Phi(t),\varphi(t),n))\equiv \Im(u(t,n))\not=0.
\end{equation}

In what follows, we will study the model based on the analysis of the behavior of its basic functions: the scale factor $a(t)$ (or $\xi(t)$), the Hubble parameter $H(t)$, the invariant cosmological acceleration $\Omega(t )$ \eqref{Omega} and invariant curvature $K(t)$ \eqref{K}, local per\-tur\-ba\-tion growth rate $\gamma(t)$ \eqref{g,o}, perturbation am\-pli\-tude growth factor $\chi(t )$ \eqref{chi} and its final value $\chi_\infty$ \eqref{chi8}.

Let us note here, so as not to return to this later, that the constant value of the Hubble parameter $H(t)=\mathrm{Const}$ corresponds to the in\-fla\-tion mode $\Omega=1$: infla\-tio\-nary expansion for $H>0$ or infla\-tio\-nary contraction for $H<0$. These two modes are mutually invertible under time inversion $t\leftrightarrow -t$.

Let us make the following remark regarding the parameters of the models under consideration: below, we will mainly consider models with small values of the scalar charge $e\leqslant 10^{-4}$, which in ordinary units corresponds to the values
\[e\lesssim 10^{-4}m^{1/2}_{pl}\sim \sqrt{10^{15} \mbox{Gev}}, \]
since even these values already lie at the level of the values of the parameters of the SU(5) field-theoretic models. Values of scalar charges on the order of $1$ would mean that scalar charges must have a gravitational nature, which would require a reformulation of the theory of gravity. For such values of the magnitude of scalar charges, according to the WKB applicability condition \eqref{WKB}, even in the case of small values of the wave number $n\lesssim1$, we can use the results of the WKB instability theory for $n\eta \gg1$, i.e., for fulfillment of the condition
\begin{equation}\label{neta>>1}
nR(t)\gg 1, \qquad \biggl(R(t)\equiv\int\limits_{t_0}^t \mathrm{e}^{-\xi}dt\biggr).
\end{equation}
\section{Numerical modeling}

\subsection{Negative values of the cosmological constant : $\Lambda<0$\label{Raz_L<0}}

Studies show (see \cite{GC_21_3}) that negative values of the cosmological constant correspond to cosmological models with a finite lifetime. At the same time, in the case of sufficiently large values of the scalar charges and the cosmological constant, the cosmological model remains stable. This corresponds, for example, to the case $\mathbf{P}=$ $[[1,1,0.001,1],$ $-0.001]$. With large negative values of the cosmological constant, among other things, the Universe has a too short lifetime. Let us therefore consider the case of sufficiently small values of the scalar charges and the cosmological constant:
\begin{equation}\label{Params1}
\mathbf{P_1}=[[1,1,10^{-5},0.1],-10^{-5}].
\end{equation}

On Fig. \ref{ignatev5} and Fig. \ref{ignatev6} the evolution of the functions $\xi(t)$ and $H(t)$ for these parameter values is shown. As can be seen from these figures, the lifetime of the cosmological model in the case of parameters \eqref{Params1} is less than 1200 Planck times.

In this case, the cosmological model is mostly in a state with an almost zero value of the Hubble parameter $H\approx0$ (Fig. \ref{ignatev6}).

\Fig{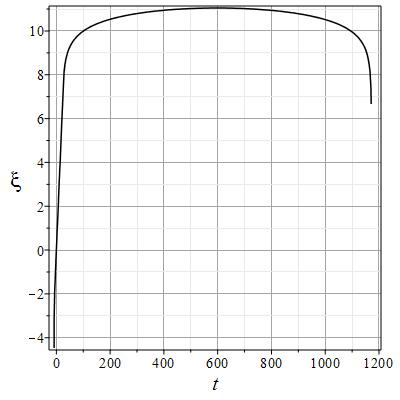}{7}{\label{ignatev5}Evolution of the scale function $\xi(t)=\ln a(t)$ in the case of parameters \eqref{Params1}.}
\Fig{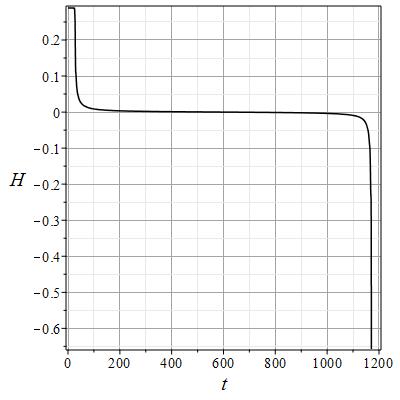}{7}{\label{ignatev6}Evolution of the Hubble parameter $H(t)$ in the case of \eqref{Params1} parameters. }

Note that the initial singularity at the parameters \eqref{Params1} in our model corresponds to the time $t_0\approx -20.12$. On Fig. \ref{ignatev7} shows the graph of the function $R(t)$ \eqref{neta>>1} for the studied model parameters. Thus, the WKB approximation \eqref{neta>>1} for $t\geqslant 0$ in the case under study takes the form:
\[1200 n\gg1,\]
which allows us to study perturbations with wave numbers $n\gtrsim 10^{-2}$ in the WKB approximation. We will not return to this issue in the future.

On Fig. \ref{ignatev8} -- \ref{ignatev9} the evolution of the perturbation frequency $\omega(t)$ and the oscillation growth increment $\gamma(t)$ are shown. As can be seen from these figures, firstly, in the case under consideration, there are 4 oscillation modes (\eqref{u_pm} and \eqref{g,o}). Secondly, the perturbation growth rate for oscillation modes $(^\pm_+)$ is strictly equal to zero $\gamma^\pm_+=0$ (in Fig. \ref{ignatev7} the lines $\gamma^\pm_+(t )$ merge). This means that the perturbation modes $(^\pm_+)$ in this case represent two pairs of undamped waves with a phase velocity \eqref{v_f}. This case corresponds to the 2nd \eqref{a^2>b>0} type of disturbances.

\Fig{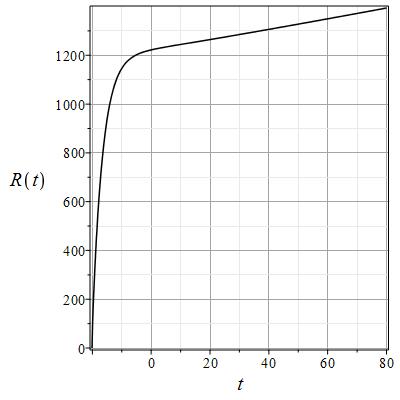}{7}{\label{ignatev7}Evolution of the Hubble parameter $H(t)$ in the case of \eqref{Params1} parameters.}
Thirdly, the oscillation mode $(^-_-)$ on the interval $[t_1,t_2]\approx[5,25]$ is unstable, while the mode $(^+_-)$ is damped on this interval. In this interval, the oscillation frequencies $\omega^\pm_-$ vanish.
Thus, on the interval $t\in[5,25]$, the perturbation modes of the $(^\pm_-)$ perturbations represent a superposition of growing and damping standing waves, i.e., the perturbations on this interval belong to the 1st type of perturbations \eqref{b>0,b>a^2}. Outside this interval, the per\-tur\-ba\-tions are pairs of non-damped waves, which corresponds to the second type of per\-tur\-ba\-tions \eqref{a^2>b>0}. In what follows, for convenience, we will refer to the type of instability corresponding to Fig. \ref{ignatev9}, \emph{narrowband instability}.

To remove possible confusion in connection with the isolation of the negative frequency part of perturbations, we recall that we must add complex conjugate quantities to the final expressions for perturbations according to the formula \eqref{Sol_perts1}, and therefore the isolation of the negative frequency part of perturbations turns out to be fictitious.
\Fig{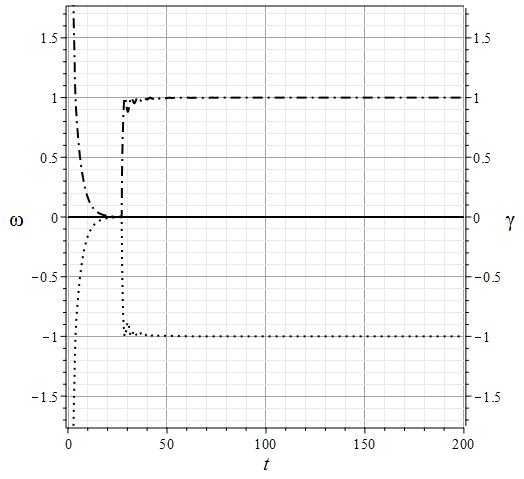}{7}{\label{ignatev8}Evolution of the oscillation frequency $\omega^\pm_+$ ($\omega^-_+$ -- dotted lines and $\omega^+_+$ -- dash-dotted lines) and the oscillation growth increment $\gamma^\pm_+ $ ($\gamma^-_+$ -- dashed and $\gamma^+_+$ -- solid lines) in case of \eqref{Params1} and $n=5$ parameters.}
\Fig{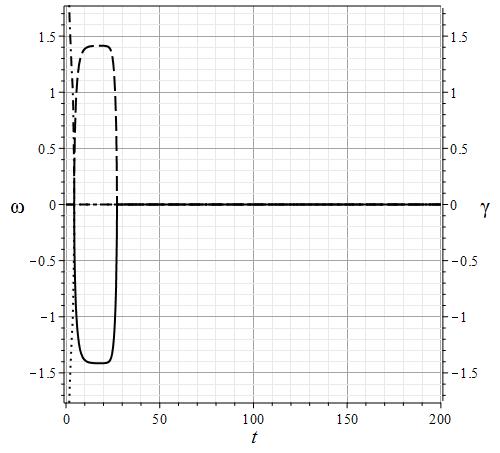}{7}{\label{ignatev9}Evolution of the oscillation frequency $\omega^\pm_-$ ($\omega^-_-$ -- dotted lines and $\omega^+_-$ -- dashed-dotted lines) and the growth rate of oscillations $\gamma^\pm_- $ ($\gamma^-_-$ are dashed and $\gamma^+_-$ are solid lines) in case of \eqref{Params1} and $n=5$ parameters.}

Let us give an example with even smaller values of the charges and the cosmological constant
\begin{equation}\label{Params1_a}
\mathbf{P_1}=[[1,1,10^{-6},0.1],-10^{-7}].
\end{equation}
In this case, for a long time, the background solution coincides with the inflationary $H=H_0\approx 0.2887$ with high accuracy. On Fig. \ref{ignatev10} the evolution of the potential of the scalar field $\Phi(t)$ for this case and the case of parameters \eqref{Params1} is shown.
\Fig{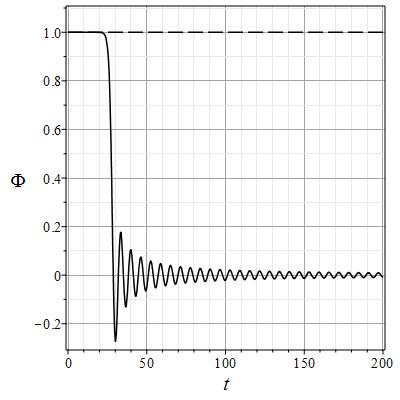}{7}{\label{ignatev10}Evolution of the scalar potential $\Phi(t)$ in the case of parameters \eqref{Params1} (solid line) and \eqref{Params1_a} (dashed line).}
This background solution corresponds to the sin\-gular point $M^+_{+1}$ with $H=H_0$, $\Phi=1$.

\Fig{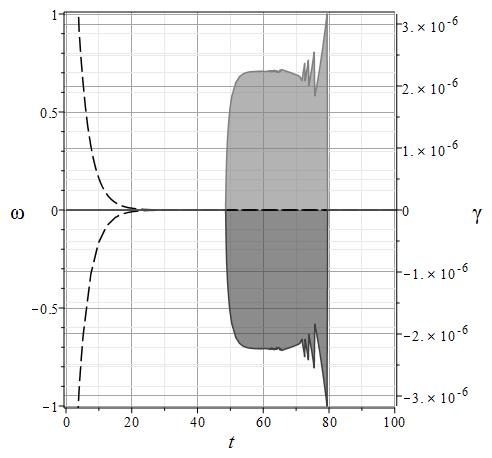}{7}{\label{ignatev11}The evolution of the oscillation frequency $\omega^\pm_+$ ($\omega^-_+$ -- dashed lines and $\omega^+_+$ -- dotted lines) and the growth rate of oscillations $\gamma^\pm_+$ ( $\gamma^-_+$ and $\gamma^+_+$ are curves enveloping light gray and dark gray areas, respectively) in case of \eqref{Params1_a} and $n=5$ parameters.}

On Fig. \ref{ignatev11} shows the evolution of the per\-tur\-ba\-tion frequency $\omega^\pm_+(t)$ and increment $\gamma^\pm_+(t)$, and Fig. \ref{ignatev12} -- evolution of per\-tur\-ba\-tion frequency $\omega^\pm_-(t)$ and increment $\gamma^\pm_-(t)$. Thus, the per\-tur\-ba\-tion modes $(^\pm_-)$ almost over the entire time interval $t\in(0.600)$ represent a pair of growing and damping standing waves with increment $\gamma\pm\approx \pm 1.43$, and the per\-tur\-ba\-tion modes $(^\pm_+)$ almost over the entire time interval $t\in(0,600)$ represent a pair of undamped delayed and advanced waves with oscillation fre\-quen\-cies decreasing with time $\omega^\pm_+(t)\to 0\ (t \to \infty)$\footnote{The burst of the amplitude increase increment on the interval $t\in(50,80)$ has a very small value $\gamma^+_-\sim 10^{-6}.$}. This means that almost on the entire interval $t\in(0,600)$ per\-tur\-ba\-tions are of the first type \eqref{b>0,b>a^2}. In what follows, for convenience, we will call the one shown in Fig. \ref{ignatev11} picture \emph{broadband instability}. Its distinguishing feature is the constancy of $\gamma(t)$ over a wide time interval. This instability arises for the per\-tur\-ba\-tion mode $(^-_-)$ in the case of a sufficiently small scalar charge $e\leqslant 10^{-5}$.
\Fig{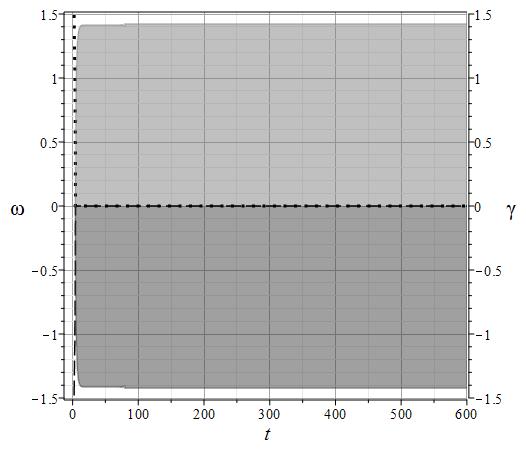}{7}{\label{ignatev12}The evolution of the oscillation frequency $\omega^\pm_-$ (dashed lines) and the growth rate of oscillations $\gamma^\pm_-$ ($\gamma^-_-$ and $\gamma^+_-$ -- curves, envelopes light gray and dark gray areas, respectively) in case of \eqref{Params1_a} and $n=5$ parameters.}

\subsection{Zero value of the cosmological constant: $\Lambda=0$\label{Raz_L=0}}
As studies \cite{GC_21_3} show, in the case of a zero value of the cosmological constant, the cosmological model \MIO\ has an infinite lifetime, the scaling function $\xi(t)$, like the scaling factor $a(t)$, is a mo\-no\-to\-ni-cally increasing function ; the Hubble parameter -- monotonically decreasing with zero asymptote $t\to+\infty$ $H(t)\to 0$. On Fig. \ref{ignatev13} shows the evolution of the function $H(t)$ for parameters
\begin{equation}\label{Params2}
\mathbf{P_2}=[[1,1,10^{-6},0.1],0].
\end{equation}
On Fig. \ref{ignatev14} shows the evolution of the perturbation frequencies $\omega^\pm_+(t)$ and the oscillation growth increment $\gamma^\pm_+(t)$, and Fig. \ref{ignatev15} -- $\omega^\pm_-(t)$ and $\gamma^\pm_-(t)$. As can be seen from the figure, this case is similar to the case with a negative cosmological constant considered in the \ref{Raz_L<0} section. In this case, $\gamma^+_-\approx 1.41$.
\Fig{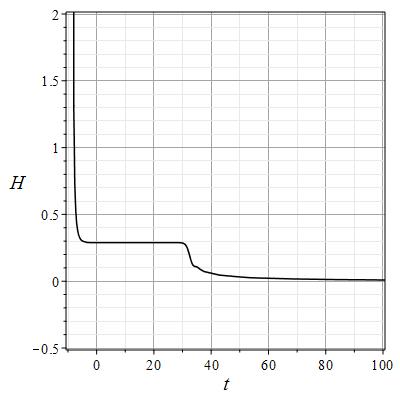}{7}{\label{ignatev13}Evolution of the Hubble parameter $H(t)$ in the case of \eqref{Params2} parameters.}

Thus, this case also corresponds to the 2nd \eqref{a^2>b>0} type of disturbances.
\Fig{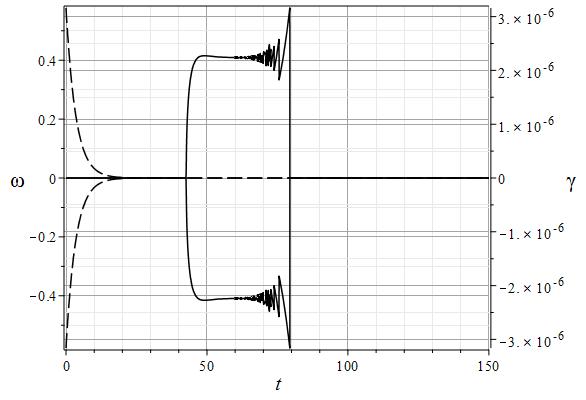}{7}{\label{ignatev14}The evolution of the oscillation frequency $\omega^\pm_+$ (dashed and dash-dotted lines) and the oscillation growth rate $\gamma^\pm_+$ (solid lines) in the case of the parameters \eqref{Params2} and $n=5$. }
\Fig{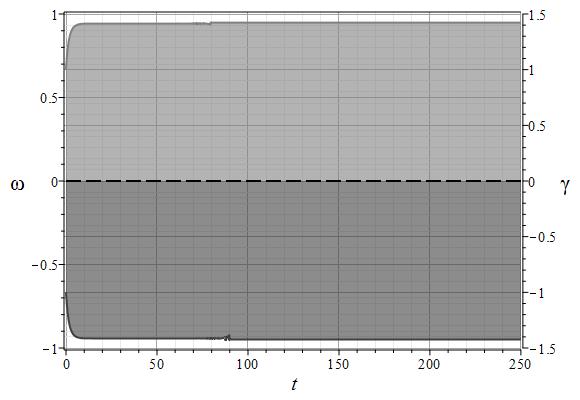}{7}{\label{ignatev15}The evolution of the oscillation frequency $\omega^\pm_-$ (dashed lines) and the growth rate of oscillations $\gamma^\pm_-$ ($\gamma^-_-$ and $\gamma^+_-$ -- curves, envelopes light gray and dark gray areas, respectively) in case of \eqref{Params2} and $n=5$ parameters.}
\subsection{Positive values of the cosmological constant\label{Raz_L>0}}
On Fig. \ref{ignatev16} shows the evolution of the function $H(t)$ for parameters
\begin{equation}\label{Params3}
\mathbf{P_3}=[[1,1,1,10^{-5},0.1],10^{-5}].
\end{equation}
In the \MIO\ model with a positive cosmological constant, a singularity appears at some time $t_0$, which should be considered the beginning of the Universe \cite{GC_21_3} in the chosen time scale. In the case of parameters \eqref{Params3} and initial conditions $\mathbf{I}=[1,0,1]$, this time corresponds to the value $t_0\approx -20.1$. The function $\xi(t)$ is monotonically increasing, and the Hubble parameter $H(t)$ is monotonically decreasing with the asymptote $t\to+\infty$ $H(t)\to H_0\approx 0.2886809080$ corresponding to the stationary point dynamical system for the Higgs vacuum scalar field \eqref{sing_point}.
The values of the Hubble parameter $H$ corresponding to singular points \eqref{sing_point} are shown in Fig. \eqref{ignatev16} with dashed lines. The asymptote $t\to+\infty$ $H(t)\to H_0\approx0.18$ corresponds to the inflationary expansion of the model.

Next, in Fig. \ref{ignatev17} graphs of the evolution of the oscillation frequency $\omega$ and the increment/decrement of the rise of oscillations $\gamma(t)$ for the oscillation modes $^\pm_+$ are presented, and Fig. \ref{ignatev18} -- for oscillation modes $^\pm_-$ in case of \eqref{Params3} parameters and wavenumber $n=5$.

\Fig{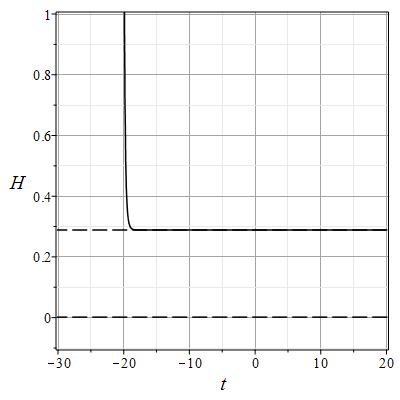}{6}{\label{ignatev16}Evolution of the Hubble parameter $H(t)$ in the case of \eqref{Params3} parameters.}
\Fig{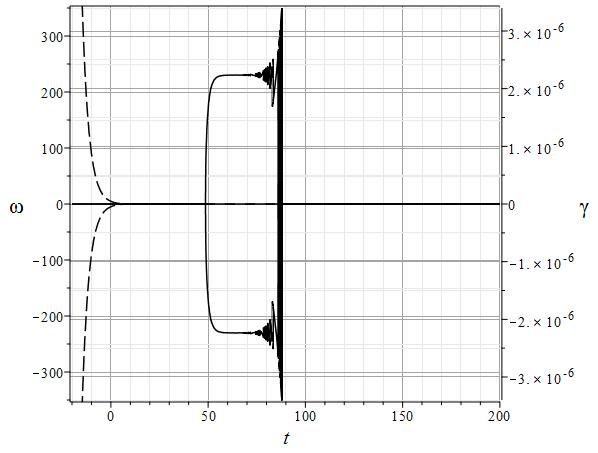}{6}{\label{ignatev17}The evolution of the oscillation frequency $\omega^\pm_+$ (dashed and dash-dotted lines) and the oscillation growth rate $\gamma^\pm_+$ (solid lines) in the case of the parameters \eqref{Params2} and $n=5$. }

Comparing the corresponding graphs in the sections \ref{Raz_L<0} ($\Lambda<0$), \ref{Raz_L=0} ($\Lambda=0$) and \ref{Raz_L>0} ($\Lambda>0 $) we see almost the same behavior of increments / \!\! decrements of the perturbation amplitude. However, this is only a superficial similarity: firstly, the fre\-quen\-cies of $\omega^\pm_+$ in these cases differ significantly: for $\Lambda<0$ $|\omega^\pm_+|\lesssim0.3$, for $ \Lambda=0$ $|\omega^\pm_+|\lesssim0.5$, for $\Lambda>0$ $|\omega^\pm_+|\lesssim300$ (Pictures [\ref{ignatev11}--\ ref{ignatev12}], [\ref{ignatev14}--\ref{ignatev15}], [\ref{ignatev17}--\ref{ignatev18}]).
 Secondly, the behavior of the function $\gamma(t)$ coincides in the case of completely different values of the parameters $e,\Lambda$: in the case of $\Lambda>0$, the behavior of $\gamma(t)$ is similar to the behavior in the case of the parameters \eqref{Params3} is achieved with parameters \eqref{Params1_a} for which the values $e,|\Lambda|$ are an order of magnitude and 2 orders of magnitude less than the corresponding parameters \eqref{Params3}. In the case of the same absolute value of the parameters \eqref{Params1} and \eqref{Params3}, the behavior of the functions $\omega$(t) and $\gamma(t)$ in these cases is fundamentally different (cf. Fig. [\ref{ignatev8} -- \ref{ignatev9}] and [\ref{ignatev17}--\ref{ignatev18}]).

\Fig{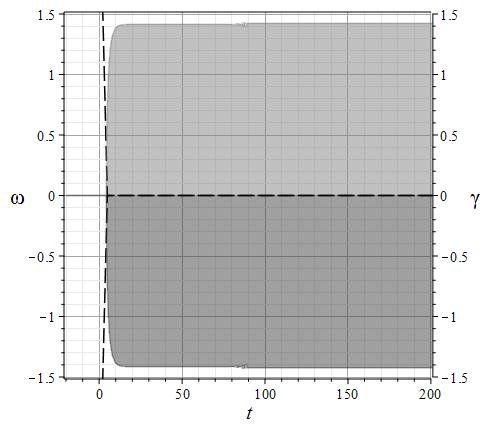}{6}{\label{ignatev18}Frequency evolution $\omega^-_-$ -- solid lines, $\omega^+_-$ -- dotted lines, and oscillation growth increment $\gamma^-_-$ -- curves bounding the light gray areas in the upper part of the graph, $\gamma^+_-$ are the curves bounding the dark gray areas in the lower part of the graph in case of \eqref{Params3} and $n=5$ parameters.}
\subsection{Influence of the value of the wave number on the growth rate\newline of oscillations}
On Fig. \ref{ignatev19} the dependence of the evolution of the growth rate of oscillations on the wave number $n=1\div 10000$ for the $(^+_-)$ mode is demonstrated.
\Fig{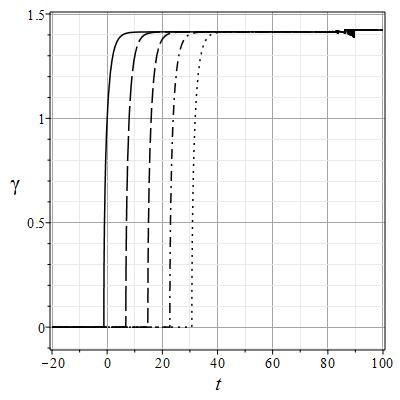}{6}{\label{ignatev19} The evolution of the oscillation growth increment $\gamma^-_-$ depending on the wave number $n$:
$n=1$ is a solid line, $n=10$ is a long dashed line, $n=100$ is a dashed line, $n=1000$ is a dashed line and $n=10000$ - - dotted line in case of \eqref{Params3} parameters.}
Thus, as the wave number $n$ increases, the beginning of the instability phase slowly shifts towards later times, while the absolute value of the growth rate $\gamma$ does not change.

\section{Large scalar charges: a type of quasi-periodic instability}
Studies have shown that for sufficiently small values of the scalar charge and the cosmological constant $e\leqslant10^{-5}$ and $\Lambda\leqslant10^{-5}$ the functions $\gamma(t)$ are practically independent of the quantities charge and cosmological constant. However, as the scalar charge increases in the $e\geqslant 10^{-5}$ region, the situation changes radically.
On Fig. \ref{ignatev20} the dependence of the evolution of the oscillation growth increment on the value of the scalar charge $e=10^{-5}\div 7\cdot10^{-5}$ for the $(^+_-)$ mode at $n=3 $. This figure clearly shows how, with an increase in the scalar charge, the broadband instability transforms into a narrowband one, which, in turn, transforms into \emph{quasiperiodic instability}. This last type of instability is characterized by alternating pauses $[t_i,t_i+\Delta t/2]$ of trapezoidal bursts of the growth rate of the disturbance amplitude $\gamma^-_-(t)$ and pauses $[t_i+\Delta t/2$, $t_i+\Delta t]$ with zero increment $\gamma^-_-(t)=0$. So, on the graph in Fig. \ref{ignatev20} $\Delta t\approx 6$.

\Fig{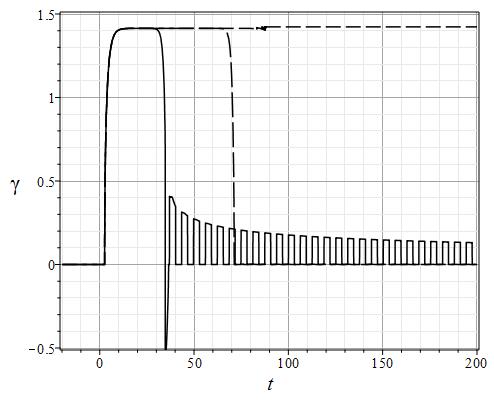}{6}{\label{ignatev20} The evolution of the growth increment of oscillations $\gamma^+_-$ depending on the value of the scalar charge:
$e=7\cdot10^{-5}$ -- solid line, $e=5\cdot10^{-5}$ -- long dashed line, $e=\cdot10^{-5}$ -- dashed line line; $n=3$, $\Lambda=10^{-5}$.}

On Fig. \ref{ignatev21} the dependence of the evolution of the growth rate of oscillations on the wave number $n=3\div 300$ for the $(^+_-)$ mode is demonstrated. Here, as in the previous cases, the magnitude of the wavenumber affects only the initial stages of the development of instability - with an increase in $n$, the beginning of the instability phase shifts to later times.
Next, in Fig. \ref{ignatev22} graphs of the evolution of the oscillation frequency $\omega$ and increment/decrement of oscillation $\gamma(t)$ for oscillation modes $^\pm_-$ in the case of parameters \eqref{Params3} and wavenumber $n= 5$.
\Fig{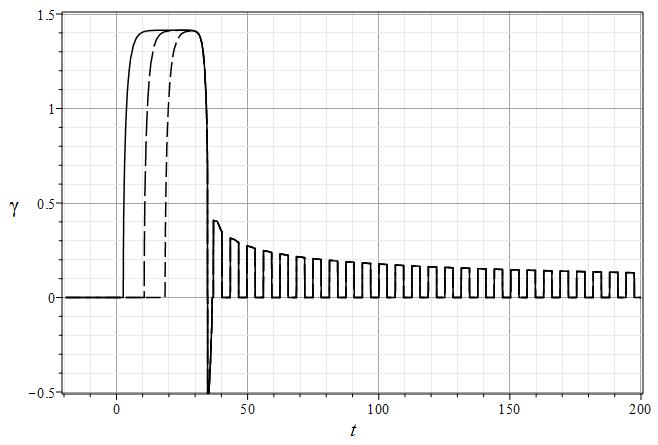}{6}{\label{ignatev21} The evolution of the oscillation growth increment $\gamma^-_-$ depending on the wave number $n$ in the case of parameters $e=7\cdot10^{-5}$, $\Lambda=10^{-5}$:
$n=3$ -- solid line, $n=30$ -- long dashed line, $n=300$ -- dashed line.}
\Fig{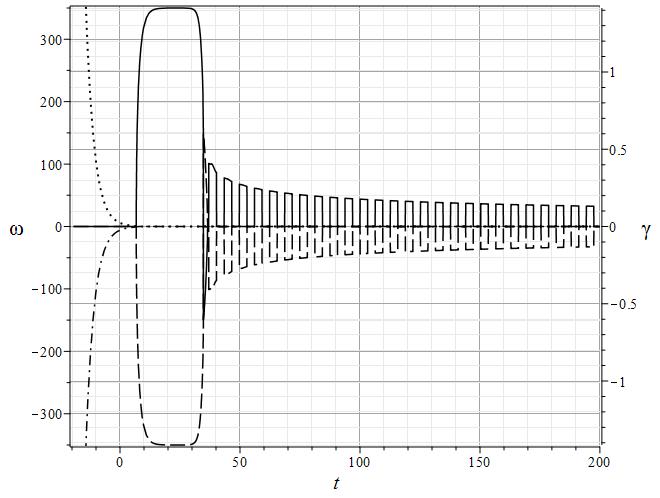}{6}{\label{ignatev22}Evolution of frequencies $\omega^-_-$ (dashed line), $\omega^+_-$ (dashed line) and growth increments $\gamma^-_-$ (solid line), $\gamma^ +_-$ (dashed line) for modes with wave number $n=10$ in case of parameters $e=7\cdot10^{-5}$, $\Lambda=10^{-5}$.}

In the transition region of sufficiently large values of the scalar charge $e\simeq 10^{-5}$ and very large values of the cosmological constant $\Lambda\simeq 0.1$, the behavior of perturbations becomes much more complicated - while maintaining the oscillatory nature of the function $\gamma(t )$ its quasi-periodicity disappears (Fig. \ref{ignatev23}). This transient type of instability can be called \emph{aperiodic instability}.

\Fig{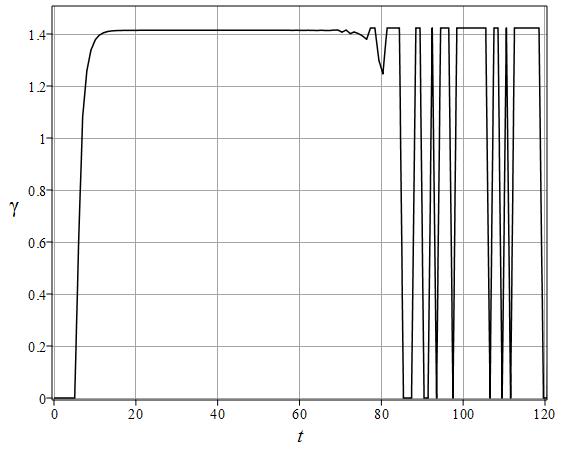}{6}{\label{ignatev23}The evolution of the oscillation growth increment $\gamma^-_-$ for modes with wavenumber $n=10$ in the case of parameters $e=10^{-5}$, $\Lambda=0.1$.}
\section{Conclusion}
Let us briefly list the main results of this part of the paper.

\stroka{A closed mathematical model is constructed for a self-consistent description of the cosmological evolution of longitudinal short-wavelength perturbations in a one-component system of degenerate scalarly charged fermions and a canonical Higgs scalar field. This model contains a system of ordinary nonlinear differential equations describing the background functions $a(t)$ and $\Phi(t)$ -- scale factor and scalar potential, as well as a system of equations describing the evolution of shortwave disturbances relative to this background.
}
\stroka{On the basis of the formulated mathematical model, expressions are obtained for the eikonal functions in the WKB approximation, which describe 4 nontrivial perturbation modes $(^\pm_\pm)$ (formulas \eqref{u_2_pm} -- \eqref{u_pm}). At the same time, the limitations of the previous works of the Author, which prevent the study of perturbations at large times and sufficiently large values of scalar charges, are removed. An analysis of the eikonal functions made it possible to identify 4 different types of perturbations, the existence of which is determined by the fundamental parameters of the model \eqref{b>0,b>a^2} -- \eqref{b<a^2,a<0}.
}
\stroka{Using the found expressions for the eikonal, the values of the frequency $\omega^\pm_\pm$ and the growth rate of the oscillation amplitude $\gamma^\pm_\pm$ are determined as functions of the fundamental parameters of the model for given background values --$\{a,\Phi \}$. With the help of numerical simulation, the fundamental possibility of the existence of regions with a non-zero growth rate of the disturbance amplitude, i.e., the possibility of a gravitational-scalar instability (Fig. \ref{ignatev1} -- \ref{ignatev4}) has been demonstrated.
}
\stroka{Four types of evolution of the $\gamma(t)$ increment are distinguished: \textbf{1.} \emph{narrow-band instability}, characterized by a short burst in the early stages of the expansion (Fig. \ref{ignatev9}, $e>10^{-5 }$, $\Lambda\gtrsim 10^{-5}$); \textbf{2.} -- \emph{broadband instability} characterized by a long phase with an approximately constant value of $\gamma^\pm_-\approx \mathrm{Const}$ (Fig. \ref{ignatev12}, \ref{ignatev15}, \ref{ignatev18} -- $e<10^{-5},\ \Lambda<10^{-5}$); \textbf{3.} -- \emph{quasi-periodic instability}, characterized by periodic alternation of pauses with a trapezoidal increment function $\gamma^-_-(t)$ with decreasing value over time and pauses with zero value $\gamma(t) =0$ (Fig. \ref{ignatev20}, \ref{ignatev22} -- $e>5\cdot10^{-5},\ \Lambda\gtrsim 10^{-5}$); \textbf{4.} -- \emph{aperiodic instability} characterized by aperiodic instability phase alternation with a constant maximum value $\gamma^-_-(t)$ (Fig. \ref{ignatev23} -- $e>10^ {-5},\ \Lambda\gg 10^{-5}$). It is shown that the value of the wave number $n$ affects the development of instability only at its earliest stages: as $n$ increases, the beginning of the unstable phase slightly shifts to later times.
}
\stroka{Since according to \eqref{chi} and \eqref{chi8} \emph{the disturbance amplitude growth factor} is determined by the area under the plot of the $\gamma(t)$ function, we can conclude that in the case of a narrow-band instability (Fig. \ ref{ignatev9}) $\chi_\infty\simeq 28$, in the case of broadband instability $\chi(t)\simeq 1.4\cdot t$ (moreover, $t$ can reach at least values of the order of $600$), in the case of quasi-periodic instability
(Fig. \ref{ignatev20}) taking into account that the duration of the ``empty'' phases is equal to half the period, we can estimate $\chi(t)$ as the sum of the area of the first burst and half the area of the curvilinear trapezium described by individual bursts $ \gamma^-_-(t)>0$: $\gamma^-_-(t)\simeq 42+0.5\cdot(t-35)$.
}

Since even in the case of narrow-band instability, the amplitude of perturbations $(^-_-)$ can increase very strongly ($\exp(28)\sim 10^{12}$ !), there is a real possibility of explaining the mechanism of formation of supermassive black holes in the early Universe using the mechanism of gravitational - scalar instability in the \MIO model, built on a one-component system of degenerate scalarly charged fermions and the classical Higgs scalar field. This issue will be explored in detail in the next part of the article.

\subsection*{Funding}

 This article was supported by the Academic Strategic Leadership Program of Kazan Federal University.


%

%
\end{document}

%% file: preamble.tex
\usepackage{amsfonts,amssymb,cite}
\usepackage{graphicx}

\def\noi{\noindent}

\newcommand{\Title}[1]{\noi {{\Large\bf #1}}\\[1ex]}

\newcommand{\Author}[2]{\noi{\bf #1}\\[2ex]\noi{\normalsize\it #2}\\}

\newcommand{\Abstract}[1]{\vskip 2mm \begin{center}
        \parbox{16.4cm}{\small\noi #1} \end{center}\medskip}
\newcommand{\foom}[1]{\protect\footnotemark[#1]}

\def\nqq{\hspace*{-2em}}

\def\Jl#1#2{#1 {\bf #2},\ }

\def\ApJ#1 {\Jl{Astroph. J.}{#1}}
\def\CQG#1 {\Jl{Class. Quantum Grav.}{#1}}
\def\DAN#1 {\Jl{Dokl. AN SSSR}{#1}}
\def\GC#1 {\Jl{Grav. Cosmol.}{#1}}
\def\GRG#1 {\Jl{Gen. Rel. Grav.}{#1}}
\def\JETF#1 {\Jl{Zh. Eksp. Teor. Fiz.}{#1}}
\def\JETP#1 {\Jl{Sov. Phys. JETP}{#1}}
\def\JHEP#1 {\Jl{JHEP}{#1}}
\def\JMP#1 {\Jl{J. Math. Phys.}{#1}}
\def\NPB#1 {\Jl{Nucl. Phys. B}{#1}}
\def\NP#1 {\Jl{Nucl. Phys.}{#1}}
\def\PLA#1 {\Jl{Phys. Lett. A}{#1}}
\def\PLB#1 {\Jl{Phys. Lett. B}{#1}}
\def\PRD#1 {\Jl{Phys. Rev. D}{#1}}
\def\PRL#1 {\Jl{Phys. Rev. Lett.}{#1}}

\def\lal{&&\nqq {}}

\def\beq{\begin{equation}}
\def\eeq{\end{equation}}
\def\bear{\begin{eqnarray}}
\def\bearr{\begin{eqnarray} \lal}
\def\ear{\end{eqnarray}}
\def\earn{\nonumber \end{eqnarray}}

